\documentclass[twocolumn,showpacs,preprintnumbers,amsmath,amssymb]{revtex4}

\usepackage{graphicx}
\usepackage{epsfig}
\usepackage{dcolumn}
\usepackage{bm}

\begin{document}

\preprint{HEP/123-qed}
\title{Quasi-free photoproduction of $\eta$-mesons off the neutron}
\author{
  I.~Jaegle$^1$,
  T.~Mertens$^1$,
  A.V. Anisovich$^{2,3}$,
  J.~C.~S.~Bacelar$^4$,
  B.~Bantes$^5$,
  O.~Bartholomy$^2$,
  D.~Bayadilov$^{2,3}$,
  R.~Beck$^2$,
  Y.A.~Beloglazov$^3$,
  R.~Castelijns$^4$,
  V.~Crede$^{2,6}$, 
  H.~Dutz$^5$,
  A.~Ehmanns$^2$,
  D.~Elsner$^5$,
  K.~Essig$^2$, 
  R.~Ewald$^5$,
  I.~Fabry$^2$,
  M.~Fuchs$^2$,
  Ch.~Funke$^2$,
  R.~Gothe$^5$,
  R.~Gregor$^7$,
  A.~B.~Gridnev$^3$,
  E.~Gutz$^2$,
  S.~H\"offgen$^5$,
  P.~Hoffmeister$^2$,
  I.~Horn$^2$,
  J.~Junkersfeld$^2$,
  H.~Kalinowsky$^2$,
  S.~Kammer$^5$
  V.~Kleber$^5$
  Frank~Klein$^5$,
  Friedrich~Klein$^5$,
  E.~Klempt$^2$,
  M.~Konrad$^5$,
  M.~Kotulla$^{1,7}$,
  B.~Krusche$^1$,
  M.~Lang$^2$,
  J.~Langheinrich$^5$,
  H.~L\"ohner$^4$,
  I.V.~Lopatin$^3$,
  J.~Lotz~$^2$,
  S.~Lugert$^7$,
  D.~Menze$^5$,
  J.G.~Messchendorp$^4$,
  V.~Metag$^7$,
  C.~Morales$^5$,
  M.~Nanova$^7$,
  V.A. Nikonov$^{2,3}$,
  D. Novinski$^{2,3}$,
  R.~Novotny$^7$,
  M.~Ostrick$^5$,
  L.M.~Pant$^7$,
  H.~van Pee$^{2,7}$,
  M.~Pfeiffer~$^7$,
  A.~Radkov$^3$,
  A. Roy$^7$,
  A.V.~Sarantsev$^{2,3}$,
  S.~Schadmand$^7$,
  C.~Schmidt$^2$,
  H.~Schmieden$^5$,
  B.~Schoch$^5$,
  S.V.~Shende$^4$,
  V. Sokhoyan$^{2}$,
  A.~S{\"u}le$^5$,
  V.V.~Sumachev$^3$,
  T.~Szczepanek$^2$,
  U.~Thoma$^{2,7}$,
  D.~Trnka$^7$,
  R. Varma$^7$,
  D.~Walther$^5$,
  Ch.~Weinheimer$^2$,
  Ch.~Wendel$^2$\\
(The CBELSA/TAPS Collaboration)
}
\affiliation{
  $^1$Department Physik, Universit\"at Basel, Switzerland\\
  $^2$\mbox{Helmholtz-Institut f\"ur Strahlen- u. Kernphysik, Universit\"at Bonn, Germany}\\
  $^3$Petersburg Nuclear Physics Institute, Gatchina, Russia\\
  $^4$KVI, Groningen, The Netherlands\\
  $^5$Physikalisches Institut, Universit\"at Bonn, Germany\\
  $^6$\mbox{Department of Physics, Florida State University, Tallahassee, USA}\\
  $^7$\mbox{II. Physikalisches Institut, Universit\"at Gie{\ss}en, Germany}\\
}
\date{\today}

\begin{abstract}
Quasi-free photoproduction of $\eta$-mesons off nucleons bound in the deuteron 
has been measured with the CBELSA/TAPS detector for incident photon energies 
up to 2.5 GeV at the Bonn ELSA accelerator. The $\eta$-mesons have been 
detected in coincidence with recoil protons and recoil neutrons, which allows 
a detailed comparison of the quasi-free $n(\gamma,\eta)n$ and $p(\gamma,\eta)p$
reactions. The excitation function for $\eta$-production off the neutron shows
a pronounced bump-like structure at $W$=1.68 GeV ($E_{\gamma}\approx$ 1 GeV), 
which is absent for the proton.
\end{abstract}
\pacs{PACS numbers: 
13.60.Le, 14.20.Gk, 14.40.Aq, 25.20.Lj
}

\maketitle

The excitation spectrum of the nucleon is closely connected to the 
properties of Quantum-Chromo Dynamics (QCD) in the low-energy,
non-perturbative regime. Lattice gauge calculations have provided results 
for the ground state properties, and recently also for some excited states 
(see e.g. Ref. \cite{Burch_06}), but the prediction of the full  
spectrum is still out of reach. The interpretation of experimental 
observations is mostly done with QCD inspired quark models. However, so far 
a comparison of the known excitation spectrum to model predictions reveals 
severe problems for all models. The ordering of some of the lowest lying 
states is not reproduced and all models predict many more states than have 
been observed. However, since the experimental data base is dominated by 
elastic pion scattering, it may be biased towards states that couple strongly 
to $\pi N$. Therefore photon induced reactions, which nowadays can 
be investigated with comparable precision as hadron induced reactions, 
have moved into the focus. 

Experiments for the study of the free proton are well developed, 
but much less effort has gone into the investigation of the neutron. 
The reason is the non-availability of free neutrons as targets. However, 
such measurements are required for the extraction of the isospin structure 
of the electromagnetic excitations. 
An excellent example is photo- and electro-production of $\eta$ mesons. 
It has been studied in detail off the free proton 
\cite{Krusche_95,Ajaka_98,Bock_98,Armstrong_99,Thompson_01,Renard_02,Dugger_02,Crede_05,Bartholomy_07,Elsner_07,Denizli_07}, 
where in the threshold region it is dominated by the excitation 
of the S$_{11}$(1535) resonance \cite{Krusche_97}. Photon beam
asymmetries \cite{Ajaka_98,Elsner_07} and angular distributions 
\cite{Krusche_95,Crede_05,Bartholomy_07}
reveal a small contributions from the D$_{13}$(1520) resonance via an
interference with the S$_{11}$. Further weak contributions of higher 
lying resonances have been suggested by detailed analyses of the data 
with different models (see e.g. \cite{Chiang_02,Anisovich_05}).
The investigation of $\eta$ photoproduction off $^2$H and $^{3,4}$He 
\cite{Krusche_95a,Hoffmann_97,Hejny_99,Weiss_01,Hejny_02,Weiss_03,Pfeiffer_04} 
has clarified the isospin structure of the S$_{11}$(1535) electromagnetic 
excitation, which is dominantly iso-vector \cite{Krusche_03} with a 
value of 2/3 for the neutron/proton cross section ratio. At higher energies, 
models predict a rise of the ratio due to other resonances. 
In the work of Chiang et al. \cite{Chiang_02} ('Eta-MAID'), 
the largest contribution comes from the D$_{15}$(1675), which has a 
strong electromagnetic coupling to the neutron \cite{PDG}. However, also
in the framework of the chiral soliton model \cite{Arndt_04} such a state is 
predicted. This is the nucleon-like (P$_{11}$) member of the  
anti-decuplet of pentaquarks. Very recently, Kuznetsov et al. 
\cite{Kuznetsov_07} reported a structure in the excitation function of $\eta$ 
production off quasi-free neutrons, which they interpreted as tentative 
evidence for a narrow resonance ($\Gamma\leq$ 30 MeV) at an excitation energy 
around 1.68 GeV. 

Here we report a measurement of angular distributions 
and total cross sections for quasi-free photoproduction of $\eta$ mesons off 
protons and neutrons bound in the deuteron for incident photon energies up to
2.5 GeV. The experiment was done at the tagged photon facility 
of the Bonn ELSA accelerator \cite{Husmann_88,Hillert_06} with the combined 
Crystal Barrel \cite{Aker_92} and TAPS \cite{Novotny_91,Gabler_94} calorimeters.
The setup is described in \cite{Elsner_07}. The liquid deuterium target of 
5.3 cm length was mounted in the center of the Crystal Barrel and surrounded 
by a three-layer scintillating fiber detector \cite{Suft_05} for charged 
particle identification. 

Photoproduction of $\eta$ mesons was studied via the 
$\eta\rightarrow 3\pi^o\rightarrow 6\gamma$ decay (the two-photon decay
was not used due to trigger restrictions). Events with at least six neutral 
hits were accepted if they could be combined to three $\pi^o$ mesons 
(invariant mass cut: 110 MeV $< m_{\gamma\gamma} <$ 160 MeV). The $\eta$-mesons
were then identified in the three-$\pi^0$ invariant mass spectrum. In this part
of the analysis, recoil nucleons were treated as missing particles. Their 
missing mass was calculated under the assumption of quasi-free meson production
on a nucleon at rest. The Fermi motion of the bound nucleons broadens the 
peaks, however it was still possible to separate single $\eta$ production from  
$\eta\pi$ final states. Recoil protons and neutrons in TAPS were identified 
with the plastic veto detectors in front of the BaF$_2$ crystals 
and a time-of-flight versus energy analysis. Protons in the barrel 
were accepted when at least two out of the three layers of the Inner-detector 
had responded within an angular difference of 10$^o$ to a hit 
in the barrel. Barrel hits were accepted as `neutral' when no layer of the 
Inner-detector had responded. A direct separation of neutrons and photons 
in the barrel was not possible. In events with seven neutral hits first six 
hits were assigned by the invariant mass analysis to the 
$\eta\rightarrow 3\pi^o\rightarrow 6\gamma$ decay-chain and the left-over hit
was taken as neutron. 

Absolute cross sections were derived from the target density (surface number 
density 0.26 b$^{-1}$), the incident photon flux, the decay branching ratio 
(31.35\%), the detection efficiency of the $\eta\rightarrow 6\gamma$ decay, 
and the detection efficiency for neutrons and protons. The photon flux
was determined by counting the deflected electrons and measuring the tagging 
efficiency (i.e. the fraction of correlated photons which pass the collimator) 
with a detector placed directly in the photon beam. Data have been taken in 
two runs with different photon fluxes (2.6 and 3.2 GeV electron 
energy, linearly polarized and unpolarized photons). The results agree within 
$\pm$10\%. They have been averaged and the systematic uncertainty of the flux 
is estimated as 10\%. The detection efficiency for the $\eta$ decay was
determined with Monte Carlo simulations (GEANT3 package \cite{Brun_86}) 
as discussed in Ref. \cite{Krusche_04} for pion production.
For neutrons and protons the detection efficiency was also simulated
(typical values: 30\% for neutrons, 90\% for protons), and 
for protons additionally determined from the analysis of $\eta$ production 
off the free proton (agreement between simulation and data better than 10\%). 
\begin{figure}[tht]
\begin{center}
\epsfig{figure=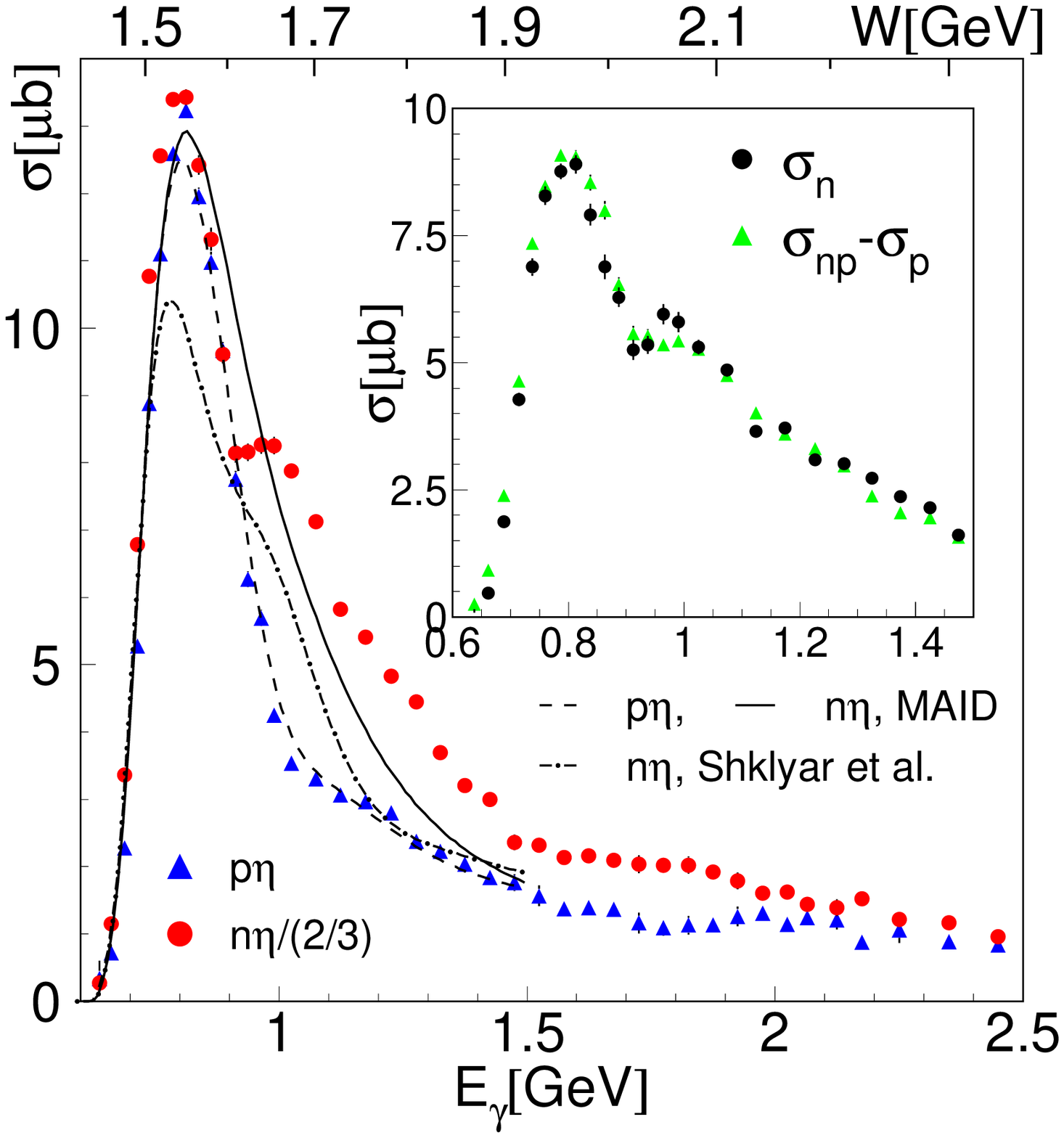,width=4.25cm}
\epsfig{figure=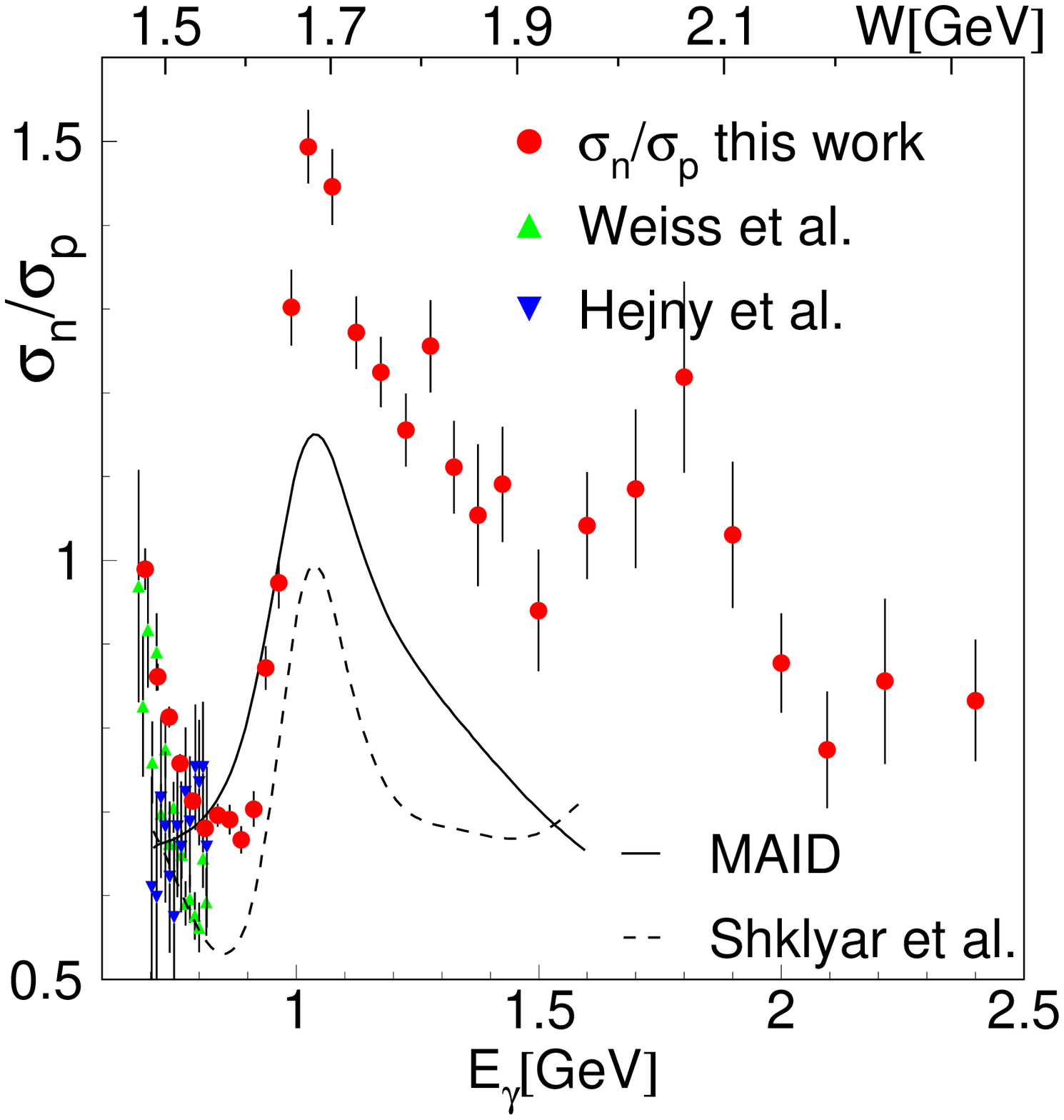,width=4.25cm}
\caption{Left hand side: quasi-free cross sections off proton 
and neutron. Curves: model predictions MAID model \cite{Chiang_02}, 
and Shklyar et al. \cite{Shklyar_07}), all folded with Fermi motion. 
Neutron data and curves are scaled up by a factor of 3/2. 
Insert: comparison of neutron cross sections from different analyses (see text).
Right hand side: neutron/proton cross section ratio. Old low energy data 
from Refs. \cite{Weiss_03,Hejny_99}
\label{fig:incl}
}
\end{center}
\end{figure}
For the reaction with coincident neutrons total systematic uncertainties from 
all sources except the photon flux have been estimated as 10\% below incident 
photon energies of 1.5 GeV, 15\% between 1.5 - 2 GeV, and below 20\% above 
2 GeV. They are below 10\% for the proton channel below 2 GeV and 15\% above. 
The only uncertainty, which does not cancel in the neutron/proton ratio,
comes from the detection of the recoil nucleons. The data were analyzed for 
$\eta$ mesons in coincidence with recoil protons ($\sigma_p$), with recoil 
neutrons ($\sigma_n$), and without any condition for recoil nucleons 
($\sigma_{np}$), including also the events without detected nucleons. 
Since coherent production can be neglected, we expect 
$\sigma_{np} = \sigma_p + \sigma_n$. Consequently, the neutron cross section 
can be determined in two independent ways, based on neutron detection 
($\sigma_n$) or on proton detection ($\sigma_{np} - \sigma_p$). The results, 
which are in excellent agreement, have been averaged and the differences can
serve as an independent estimate for the systematic uncertainties. 

\begin{figure}[th]
\begin{center}
\epsfig{figure=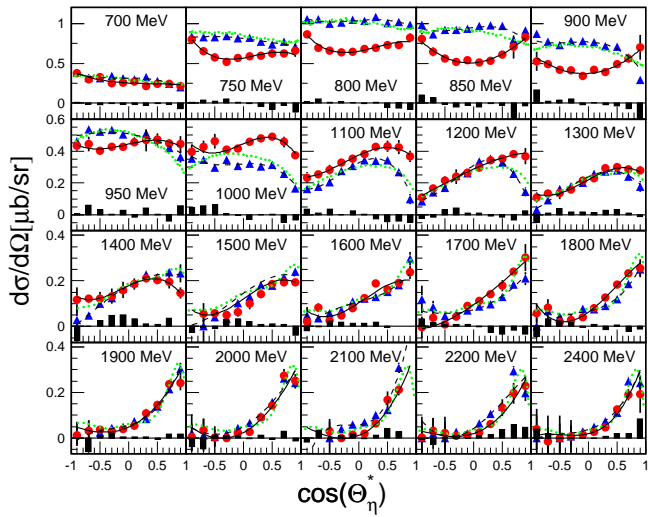,width=8.5cm}
\caption{Angular distributions: Blue triangles:  quasi-free proton, 
red circles: quasi-free neutron. Black solid (dashed) lines: 
fit of neutron (proton) data (see text), bar charts: half difference
of $\sigma_n$ and ($\sigma_{np} - \sigma_p$). Green dotted curves: free 
proton results \cite{Crede_05} folded with Fermi motion.
\label{fig:angdis}
}
\end{center}
\end{figure}

The total cross sections, which have been obtained by integration of the 
angular distributions, are shown in Fig.~\ref{fig:incl}. 
The data for the neutron show a bump-like structure around photon energies
of 1 GeV, which is not seen for the proton. This energy corresponds to a
photon - nucleon cm-energy $W\approx$1.68 GeV 
($W^2=2E_{\gamma} m_N + m_N^2$; $m_N$: nucleon mass).
The structure is particularly prominent in the neutron/proton cross section
ratio shown at the right hand side of Fig. \ref{fig:incl}. At low incident 
photon energies the ratios are in good agreement with previous results 
measured off quasi-free nucleons bound in the deuteron \cite{Weiss_03} or in 
$^4$He \cite{Hejny_99}, which have, however a larger systematic uncertainty
than the present data since the geometrical acceptance for the recoil nucleons
was limited. The data are compared in Fig.~\ref{fig:incl} to model 
calculations by Chiang et al. ('Eta-MAID') \cite{Chiang_02} and Shklyar et al. 
\cite{Shklyar_07}. Their results have been folded with 
the momentum distribution of the bound nucleons \cite{Lacombe_81} as discussed 
in \cite{Krusche_95a}. Both models show a peak-like structure in the cross 
section ratio around 1 GeV, although less pronounced than in the data.
However, the responsible mechanisms in the models are quite different.
In the MAID model \cite{Chiang_02}, the bump is mostly due to the contribution 
of the D$_{15}$(1675). The model uses a large $N\eta$-decay branching ratio 
(17\%) for this state (PDG quotes 0$\pm$1\%). Shklyar, Lenske, and Mosel 
\cite{Shklyar_07} try to explain the data with coupled channel effects 
involving the S$_{11}$(1535), S$_{11}$(1650), and P$_{11}$(1710) resonances.

\begin{figure}[th]
\begin{center}
\epsfig{figure=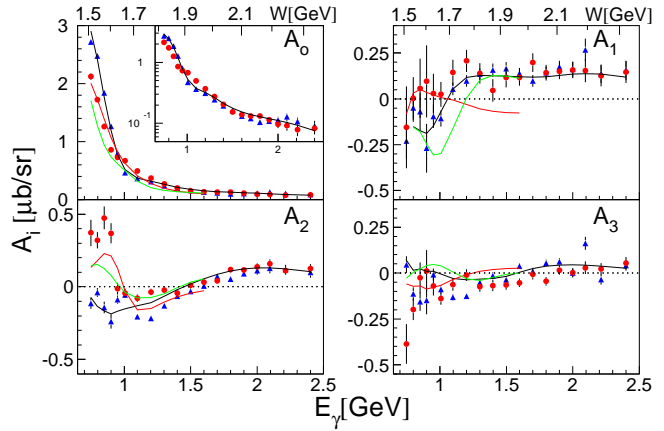,width=8.5cm}
\caption{Legendre coefficients of angular distributions. 
Red circles: quasi-free neutron, blue triangles: quasi-free proton, 
Curves: free proton data \cite{Crede_05} (black), 
MAID neutron \cite{Chiang_02} (red), Shklyar et al. neutron \cite{Shklyar_07} 
(green), all folded with Fermi motion. Insert: logarithmic scale for $A_0$.
\label{fig:coeff}
}
\end{center}
\end{figure}

The angular distributions in the cm system of the incident photon and a 
nucleon at rest (see \cite{Krusche_95a} for details) are shown in 
Fig.~\ref{fig:angdis}. 
The results for the quasi-free proton are in agreement with the free proton 
distributions folded with Fermi motion. The distributions have been fitted 
with Legendre polynomials $P_i(\mbox{cos}(\Theta^{\star}_{\eta}))$, related 
to the contributing partial waves \cite{Krusche_03}:
\begin{equation}
\frac{d\sigma}{d\Omega} = \frac{q_{\eta}^{\star}}{k_{\gamma}^{\star}}
\sum_{i} A_iP_i(\mbox{cos}(\Theta^{\star}_{\eta}))
\end{equation}
where the $A_i$ are expansion coefficients. The phase-space factor 
${q_{\eta}^{\star}}/{k_{\gamma}^{\star}}$ is also evaluated for the above cm
system. The results for $A_0,...,A_4$ are shown in Fig.~\ref{fig:coeff}. 
In the region of the dominant S$_{11}$(1535), the $s$-wave contributions 
($A_{0}$) reflect the ratio of the helicity couplings of this
resonance \cite{Krusche_03}, and the shape of the angular distributions
reflects the interference with the D$_{13}$(1520) (see Ref. \cite{Weiss_03}).
In an approximation taking into account only contributions from these two
resonances \cite{Weiss_03}, the $A_2$ coefficient is directly proportional
to the product of their helicity-1/2 couplings. This implies a positive sign 
for $A^n_2$, a negative sign for $A^p_2$, and a larger absolute value of 
$A^n_2$, all in agreement with the data. 
In this region, the angular distributions for proton and neutron are 
well described by the MAID model \cite{Chiang_02}, which qualitatively
reproduces the peak in the $A_2$ coefficient due to the S$_{11}$ - D$_{13}$
interference.
The angular distributions for protons and neutrons are similar
for high incident photon energies, where diffractive $t$-channel processes 
make a large contribution \cite{Crede_05,Bartholomy_07}. 
In the most interesting region around $E_{\gamma}\approx$ 1 GeV, the shapes 
of the angular distributions for the neutron and the proton change rapidly, 
and apart from $A_0$, their Legendre coefficients become similar.
For the proton, the $A_1$ coefficient rises sharply from negative to
positive values. Denizli et al. \cite{Denizli_07} have recently shown
in electroproduction that this feature survives to larger $Q^2$ values, and
interpreted it as an interference between S$_{11}$ and P-resonances.
However, so far there is no conclusive evidence if the structure observed 
in the neutron excitation function is related to P-resonances.
Around $E_{\gamma}\approx$1 GeV, the data are in better agreement with the 
MAID model than with the results from Shklyar et al. \cite{Shklyar_07}, 
which due to a strong S$_{11}$ - P$_{11}$ interference develop large negative 
$A_1$ coefficients. However, at somewhat higher energies, the picture changes 
completely. Altogether, the behavior of $A_1$, reflecting the S - P 
interference seems to be least well reproduced by the models.
A partial wave analysis in the framework of the Bonn-Gatchina model 
\cite{Anisovich_05}, which will be published elsewhere, can reproduce the 
bump-structure and the angular distributions alternatively with either 
interference effects in the S$_{11}$ sector, or with an additional P$_{11}$ 
state. Consequently, so far no definitive conclusion can be drawn. 

\begin{figure}[t]
\begin{center}
\epsfig{figure=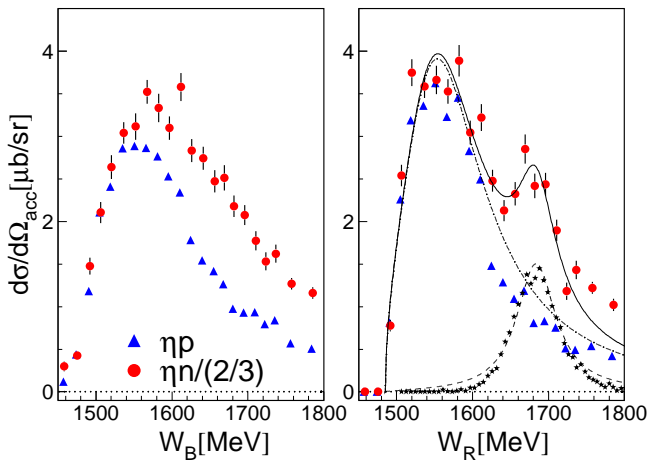,width=8.5cm}
\caption{Proton and neutron excitation functions for 
$\mbox{cos}{(\Theta_{\eta}^{\star}}) < -0.1$ without (invariant mass $W_B$ from 
incident photon energy) and with (invariant mass $W_R$ from neutron - $\eta$
4-vectors) event-by-event correction of Fermi motion. Curves: solid: full fit, 
dash-dotted: BW-curve of S$_{11}$(1535), dashed: BW-curve for second structure.
Stars: response for a $\delta$-function due to finite energy resolution.
Neutron data scaled by factor 3/2. 
\label{fig:total}
}
\end{center}
\end{figure}

Due to the Fermi smearing, it is difficult to discriminate between scenarios 
with very narrow states and broader resonances. Fix, Tiator, and 
Polyakov \cite{Fix_07} find comparable results for the MAID model with strong 
D$_{15}$ contribution and for a model with a P$_{11}$ as narrow as 10 - 30 MeV.
In principle, Fermi motion can be corrected event-by-event when energy and 
momentum of the recoil nucleons are known. Instead of using the total cm 
energy $W_B$ deduced from the incident photon energy, it can be reconstructed 
from the four-vectors of the $\eta$ meson and the recoil nucleon ($W_R$). 
Since the energy of the neutrons is measured by time-of-flight, only neutrons 
in TAPS, which correspond to $\eta$ mesons with 
cos$(\Theta_{\eta}^{\star}) < -0.1$ can be used. 
The results are summarized in Fig. \ref{fig:total}. For proton and neutron, 
the correction leads to the expected narrower peak for the S$_{11}$(1535). 
However, for the neutron also a narrow structure around $W\approx$1.7 GeV 
appears. The neutron data were fitted with the sum of two Breit-Wigner (BW) 
curves corresponding to the S$_{11}$(1535) \cite{Krusche_97} and the structure 
around 1.7 GeV. The parameters for the S$_{11}$ 
(position: 1566 MeV, width: 162 MeV) are similar to a fit of the free
proton data (1540 MeV, 162 MeV). The position of the second structure 
($W$=1683 MeV) is in agreement with the result of the GRAAL experiment 
($W\approx$ 1.68 GeV, \cite{Kuznetsov_07}). The fitted width of this structure 
is (60$\pm$20) MeV, however, this is only an upper limit, since it is broadened
by the time-of-flight resolution. Even the simulation of a $\delta$-function 
at the peak position results in a similar line-shape (see Fig.~\ref{fig:total}),
so that no lower limit of the width can be deduced. 

In summary, precise data have been measured for quasi-free photoproduction of
$\eta$-mesons off nucleons bound in the deuteron. The results for the quasi-free
proton are in excellent agreement with free-proton data, folded with the 
momentum distribution of the bound nucleons, which demonstrates the validity of
the participant - spectator approach. The data in the excitation range of the 
S$_{11}$(1535) resonance confirms the $(A_{1/2}^n/A_{1/2}^p)^2\approx 2/3$ 
ratio for this resonance and also the opposite sign of the S$_{11}$-D$_{13}$ 
interference term for proton and neutron.
At incident photon energies around 1 GeV, corresponding to $W$=1680 MeV a
pronounced bump-like structure is found in the quasi-free neutron excitation 
function, which does not exist for the proton. For the width of this structure 
in the excitation function of the free neutron only an upper limit of 
$\approx$60 MeV could be determined. Since such a structure 
can be reproduced by models based on different mechanisms, so far no final 
conclusion about its nature can be drawn.

\acknowledgments
We wish to acknowledge the outstanding support of the accelerator group 
and operators of ELSA. 
This work was supported by Schweizerischer Nationalfonds and 
Deutsche Forschungsgemeinschaft (SFB/TR-16.)

\end{document}